\documentclass[twocolumn,showpacs,amsmath,amssymb,aps,prl]{revtex4}

\pdfoutput=1

\usepackage{graphicx}

\usepackage{amsmath}

\def\e{\epsilon}
\def\t0{\tau_0}

\newcommand{\be}{\begin{equation}}
\newcommand{\ee}{\end{equation}}
\newcommand{\bea}{\begin{eqnarray}}
\newcommand{\eea}{\end{eqnarray}}

\newcommand{\tr}[2]{\mathrm{Tr}_{#1}\left[#2\right]}

\begin{document}

\title{The entanglement entropy of  one-dimensional gases} 

\author{
Pasquale Calabrese, Mihail Mintchev and Ettore Vicari
}
 \affiliation{Dipartimento di Fisica dell'Universit\`a di Pisa and INFN, Pisa, Italy}

\date{\today}

\begin{abstract}

We introduce a systematic framework to calculate the bipartite
entanglement entropy of a spatial subsystem in a
one-dimensional quantum gas which can be mapped into a noninteracting
fermion system. To show the wide range of applicability of the
proposed formalism, we use it for the calculation of the entanglement
in the eigenstates of periodic systems, in a gas confined by
boundaries or external potentials, in junctions of quantum wires and
in a time-dependent parabolic potential.

\end{abstract}
\maketitle

Entanglement is a fundamental phenomenon of quantum mechanics.  Much
theoretical work has focused on the entanglement properties of quantum
many-body systems, showing their importance to characterize the
many-body dynamics~\cite{rev}.  In particular, lots of studies have
been devoted to quantify the highly nontrivial connections between
different parts of an extended quantum system, by computing von
Neumann or R\'enyi entanglement entropies of the reduced density
matrix $\rho_A$ of a subsystem $A$.  The most remarkable result is the
universal behavior at 1D quantum critical points, determined by the
central charge of the underlying conformal field theory (CFT)
\cite{holzhey,vidalent,cc-04,cc-rev}.  For a partition of an infinite
1D system into a finite piece $A$ of length $\ell$ and the remainder,
the entanglement entropy for $\ell$ much larger than the
short-distance cutoff $a$ is \be S_{1}
\equiv-\tr{}{\rho_A\ln\rho_A}=\frac{c}3 \ln \frac{\ell}a +O(1)\,,
\label{criticalent}
\ee where $c$ is the central charge.  This result has been confirmed
in many spin-chains and in 1D itinerant systems on the lattice~\cite{rev}.  
These studies have allowed a deeper understanding of 1D
simulation algorithms based on the so-called matrix product states~\cite{mps}.  
However, the same result must be valid also in systems in
continuous space (when the UV cutoff is properly imposed). Apart from
the interest to describe trapped 1D gases experimentally realized with
cold atoms, the entanglement of continuous models is also instrumental
to develop 1D tensor network algorithms for gases, as the one proposed
in~\cite{vc-10}.  Despite of this fundamental interest, almost no
effort (with the exception of Refs.~\cite{Klich-06,bk11} and the orbital
partitioning in quantum Hall states~\cite{QH}) has been devoted 
to the spatial entanglement of gas models (that is distinguished
from the particle partitioning~\cite{mzs-09}).

In this Letter we present a systematic framework to tackle free gases
in any external conditions for an arbitrarily large number of
particles.  The most general result of this investigation is that,
when dealing with a finite number of particles $N$, the 1D
entanglement entropy grows like $\ln N$, with a prefactor that again
is given by the central charge.  In this formulation $N$ acts as an
explicit UV cutoff, representing a concrete alternative to the
lattice.  While the $\ln N$ behavior could have been predicted also on
the basis of scaling arguments, its validity in many different
physical situations is not obvious a priori.  Furthermore, being the
method very general, it allows us to derive exact predictions that, in
some cases, were not known from the lattice and/or CFT arguments.  We
apply our framework only to gases of spinless fermions, but its
validity is more general.  It is indeed straightforward to include
spin degrees of freedom. Furthermore, the 1D Bose gas in the limit of
strong short-ranged repulsive interaction (corresponding to a gas of
impenetrable bosons, describing also the dilute limit of the
finite-strength model~\cite{LL-63}) can be mapped to spinless
fermions, so that their entanglement entropies coincide.

In the following we first present the general framework based on a
mapping from a continuous Fredholm determinant into a standard one of
dimension $N$.  We then show the power of the method applying it to
several physical situations, including wires junctions and 
non-equilibrium problems that are not yet solved on the lattice.

{\it The method}.  Let us consider a system of $N$ non-interacting
spinless fermions with discrete one-particle energy spectrum.  As well
known, the many body wave functions $\Psi(x_1,...,x_N)$ can be built
from the one-particle eigenstates via $\Psi(x_1,...,x_N)={\rm det}
[\phi_k(x_n)]/\sqrt{N!}$, where the normalized wave functions
$\phi_k(x)$ represent the occupied single-particle energy levels. The
ground state is obtained by filling the lowest $N$ energy levels.
Thus, the ground-state two-point correlator reads
\begin{equation}
{\mathbb C}(x,y) \equiv \langle c^\dagger(x) c(y) \rangle = 
\sum_{k=1}^{N} \phi^*_k(x) \phi_k(y)\,, 
\label{cxy}
\end{equation} 
where $c(x)$ is the fermionic annihilation operator and the one-particle
eigenfunctions $\phi_k(x)$ are ordered according to their energies.

We want to compute the bipartite entanglement entropy of a space
interval $A$, extending from $x_1$ to $x_2$, in this fermion gas. For
this purpose, we introduce the {\it Fredholm determinant}
\begin{equation}
D_A(\lambda) =
{\rm det}\left[{\lambda {\mathbb I}-{{\mathbb C}}}\right] \,,
\label{dl}
\end{equation}
where the {\it continuous} matrix ${\mathbb C}$ and the identity
  ${\mathbb I}$ are restricted to the part at hand, i.e., from $x_1$
  to $x_2$.  Then, calculations based on the Wick theorem~\cite{ep-rev}, 
and the integral representation of Ref.~\cite{jk-04}
  for the R\`enyi entanglement entropy, allow us to write
\begin{equation}
S_\alpha(x_1,x_2) \equiv \frac{\ln {\rm Tr}\rho_A^\alpha}{1-\alpha} = 
 \oint \frac{d \lambda}{2\pi i}\, e_\alpha(\lambda) 
\frac{d \ln D_A(\lambda)}{d\lambda},
\label{snx}
\end{equation}
where the integration contour encircles the segment $[0,1]$, and
\begin{equation}
e_\alpha(\lambda) = {1\over 1-\alpha} 
\ln \left[{\lambda}^\alpha
+\left({1-\lambda}\right)^\alpha\right]\,.
\label{enx}
\end{equation}
For $\alpha\to 1$ it reduces to the von Neumann definition
(\ref{criticalent}).  The integral representation (\ref{snx}) has been
already derived and used in the context of discrete chain models~\cite{jk-04}, 
thus involving the determinant of a standard matrix with
the lattice sites as indeces.

The Fredholm determinant is turned into a standard one by introducing
the $N\times N$ {\em overlap} matrix ${\mathbb A}$ (also considered in
Ref.~\cite{Klich-06}) with elements
\begin{equation}
{\mathbb A}_{nm} =  \int_{x_1}^{x_2} dz\, \phi_n^*(z) \phi_m(z),
\qquad n,m=1,...,N,
\label{aiodef}
\end{equation}
such that ${\rm Tr}\,{\mathbb C}^k= {\rm Tr}\,{\mathbb A}^k$ for any $k$,
thus
\be
\ln {D}_A(\lambda) =
-\sum_{k=1}^\infty { {\rm Tr}  {\mathbb C}^k \over k \lambda^k }=
-\sum_{k=1}^\infty { {\rm Tr} {\mathbb A}^k \over k \lambda^k }
= 
\sum_{m=1}^N \ln(\lambda-a_m)\,, 
\label{logdetn}
\ee where $a_m$ are the eigenvalues of ${\mathbb A}$
[we use the standard
relation  for Fredholm determinants
$ \ln {\rm det}[{\mathbb I}-z
{\mathbb M}] = -\sum_{k=1}^\infty (z^k {\rm Tr} \ {\mathbb M}^k)/ k$ 
and we drop off a term $\propto\ln\lambda$ giving vanishing 
contribution in Eq.~(\ref{snx})].
Inserting it into the integral (\ref{snx}), we obtain
\begin{equation}
S_\alpha(x_1,x_2) =  \oint  \frac{d \lambda}{2\pi i}
\sum_{m=1}^N {e_\alpha(\lambda)\over  \lambda  - a_m}  = 
\sum_{m=1}^N e_\alpha(a_m),
\label{snx2n}
\end{equation}
as a consequence of the residue theorem.

The matrix ${\mathbb A}$ is easily obtained for any non-interacting
model from the one-particle wave functions, as the definition
(\ref{aiodef}) shows.  Calculating the entanglement entropies is then
reduced to an $N\times N$ eigenvalue problem that can be easily solved
numerically and in some instances even analytically, as we are going
to show.  Details of the calculations will be reported elsewhere.

\begin{figure}[t]
\includegraphics[width=0.49\textwidth]{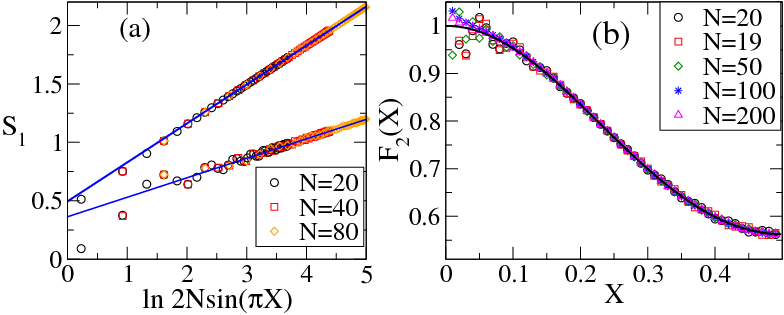}
\caption{ (a) Entanglement entropy $S_1$ for periodic (uppermost set
of data) and Dirichlet (bottom) BC compared with the asymptotic
results for different values of $N$ $(X=\ell/L)$.  (b) The function
$F_2(X)$ for the finite-size scaling behavior of the particle-hole
excited state vs the CFT prediction~\cite{abs-11}.}
\label{fig:1}
\end{figure}

{\it Ground state of a periodic system}.  In a system of length $L$
with periodic boundary conditions (BC), the one-particle
wave-functions are $\phi_{k}(x)=e^{2\pi i k x/L}/\sqrt{L}$ with
wave numbers $k\in{\mathbb Z}$.  In the ground state of the
fermion gas, filling the $N$ $k$-modes with lowest energies, the
matrix ${\mathbb A}$ associated with a segment of length $\ell$ 
is  \be {\mathbb A}_{nm}= \frac{\sin \pi(n-m)X}{\pi (n-m)},
\qquad X\equiv \ell/L,
\label{anmper}
\ee $n,m=1,...,N$.
By inserting the $N$ eigenvalues of ${\mathbb A}$ into
Eq.~(\ref{snx2n}), we obtain the entanglement entropy in a system of
$N$ particles.  Furthermore, since $\ln D_A=\ln \det {\mathbb G}$, with
${\mathbb G}\equiv \lambda {\mathbb I}-{\mathbb A}$ an $N\times N$
Toeplitz matrix, we can use the Fisher-Hartwig conjecture~\cite{fh}
to rigorously infer that
\be
S_{\alpha} =\frac16(1+\alpha^{-1})\ln(2N\sin\pi X)+ b_\alpha +
O(N^{-{2\over\alpha}}),
\label{FHres}
\ee where $b_\alpha$ and the leading $O(N^{-{2\over\alpha}})$
corrections can also be computed analytically~\cite{footnotesalpha}.
Fig.~\ref{fig:1} shows a comparison with exact finite-$N$
calculations, for $\alpha=1$.  Eq.~(\ref{FHres}) agrees with the CFT
prediction for finite systems~\cite{cc-04}, obtained from
Eq.~(\ref{criticalent}) replacing $\ell$ with the chord length $L/\pi
\sin(\pi X)$.  Thus, Eq.~(\ref{FHres}) represents the first explicit
analytic confirmation of this CFT prediction.

{\it Systems with hard-wall potential}.  We now consider a gas of
spinless fermions confined in the interval $[0,L]$ by a hard-wall
potential (i.e., the gas density vanishes for $x\notin [0,L]$). We
consider a bipartition starting from the boundary, i.e., $A=[0,\ell]$,
but the general case can also be treated.  Some algebra similar to before
leads to \be {\mathbb A_{nm}} = {\mathbb
B}_{nm}(X)\equiv{\sin[\pi(n-m)X] \over \pi(n-m)} - {\sin[\pi(n+m)X]
\over \pi(n+m)}.
\label{Snm}
\ee 
Using a recent generalization of
the Fisher-Hartwig conjecture to Toeplitz+Henkel 
matrices~\cite{idk-09,footnotesalpha}, 
we obtain 
\be
S_{\alpha} =\frac1{12}(1+\alpha^{-1})\ln(4N\sin\pi X)+ \frac{b_\alpha}2 
+ O(N^{-{1\over\alpha}}).
\label{FHres2}
\ee A comparison of the finite-$N$ results with Eq.~(\ref{FHres2})
is shown in Fig.~\ref{fig:1}.  The leading oscillating
$O(N^{-{1\over\alpha}})$ corrections can also be computed.  They
correspond to the $O(\ell^{-{1\over\alpha}})$ corrections found within
CFT~\cite{cc-10,fc-11}.

{\it Excited states} can be easily treated in this formalism by
summing in Eq.~(\ref{cxy}) over the occupied one-particle states
representing the excited states.  As an example, let us consider the
particle-hole excitation in a periodic system obtained by moving one
particle from the highest occupied level to the first available one.
The corresponding $N\times N$ matrix ${\mathbb A}$ has the first $n-1$
raws and columns identical to the ground state, and the last
different. Note that it ceases to be a Toeplitz matrix. In
Fig.~\ref{fig:1} we compare the scaling function
$F_2(X)\equiv\exp[{S_2- 1/4 \ln (2N\sin\pi X)-b_2}]$ for several
values of $N$ with the corresponding CFT prediction~\cite{abs-11}
$F_2(X)= [7+\cos(2\pi X)]^2/64$. The CFT curve is clearly approached
in the large-$N$ limit.

\begin{figure}[t]
\includegraphics[width=0.19\textwidth]{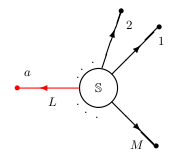}
\includegraphics[width=0.27\textwidth]{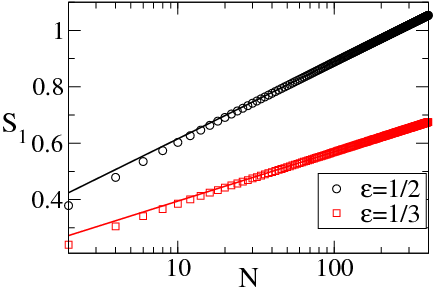}
\caption{Left: $M$ wires interacting through the scattering matrix
  ${\mathbb S}$.  Right: Entanglement entropy $S_1$ for a junction of
  two wires and for different values of the transmission coefficient $s$.
  The full lines correspond to the asymptotic behavior (\ref{asbehb}) with
  ${\cal C}_1(s)$ given by the limit $\alpha\to 1$ of Eq.~(\ref{ccalal}).}
\label{fig:jun}
\end{figure}

{\it Star graphs or wire-junctions}. A novel application of the
outlined method consists in determining the entanglement at a junction
of quantum wires (called also star graph).  Networks of quantum wires
with junctions attracted recently a lot of attention~\cite{w1}, mainly
because of their possible applications in nanocircuits.  A star graph
consists of $M$ wires joining in a single point (called vertex).  We
consider wires whose bulk is described by non-interacting spinless
fermions.  The only interaction is localized at the vertex and it is
encoded in a $M\times M$ scattering matrix ${\mathbb S}$~\cite{w1,w2},
as pictorially showed in Fig.~\ref{fig:jun}.  
We consider wires of finite length $L$ with hard-wall 
boundary conditions at their ends. Scale invariant
junctions can be constructed and classified~\cite{w1,w2}: they are
either isolated points or families with free parameters. 

Let us first consider the case of two wires, i.e., a gas of free
fermions with a localized impurity at the vertex.  The allowed scale
invariant conditions at the vertex are~\cite{w2}: the two trivial ones
(Dirichlet and Neumann) which disconnect the two wires and give no
entanglement, and the one-parameter family described by the scattering
matrix
\begin{equation} 
{\mathbb S}(\epsilon) = 
{1\over 1+\epsilon^2}\left(\begin{array}{cc}-1+\epsilon^2 & 2\epsilon\\ 
2\epsilon & 1-\epsilon^2  
\\ \end{array} \right)\, .
\label{n21}
\end{equation} 
(A phase can be added, but it does not enter any measurable
quantity and will be omitted.)  Note that $\e=1$ corresponds to full
transmission, i.e., no impurity.

In order to derive the ground-state entanglement entropy of one wire,
we compute the matrix ${\mathbb A}$ in Eq.~(\ref{aiodef}) by
diagonalizing the one-particle problem. We obtain
\begin{eqnarray}
&&{\mathbb A}_{nm} = {2\epsilon\over 1+\epsilon^2}{\mathbb B}_{nm}(1/2) 
\quad {\rm for}\;n\ne m,
\label{bmn2}\\
&&{\mathbb A}_{nn} = {1\over 1+\epsilon^2} \;\; {\rm for \; odd\;}n,\quad 
{\mathbb A}_{nn} = {\epsilon^2 \over 1+\epsilon^2} \;\; {\rm for \; even\;}\,n,
\nonumber
\end{eqnarray}
where ${\mathbb B}_{nm}$ is defined in Eq.~(\ref{Snm}). Then, we
determine the wire entanglement entropy $S_\alpha(s;N)$ as a function
of the transmission coefficient
\begin{equation}
s \equiv {2\epsilon\over 1+\epsilon^2}.
\label{sdef}
\end{equation}
We observe a logarithmic growth
\begin{equation}
S_\alpha(s;N) = {\cal C}_\alpha(s) \ln N + O(1)\,,
\label{asbehb}
\end{equation}
with a prefactor ${\cal C}_\alpha(s)$ that depends on $s$ and not only
on the central charge, see Fig.~\ref{fig:jun}.  Assuming universality
in terms of $s$, we can exploit the results of Ref.~\cite{EP-10} for
the lattice Ising and XX models with a defect, to infer
\begin{equation}
{\cal C}_\alpha(s) = {2\over \pi^2 (1-\alpha)}  \int_0^\infty dx  
\ln\left[  1 + e^{-2\alpha \omega(s)}  \over 
(1 + e^{-2\omega(s)})^\alpha \right],
\label{ccalal}
\end{equation}
where $\omega(s) = {\rm acosh}\left[{({\rm cosh}x)/s}\right]$.  The
perfect agreement, see Fig.~\ref{fig:jun}, with the large-$N$ behavior
of our numerical evaluations nicely confirms the universality
conjecture.

The case of a general number of wires $M$ can be analogously treated. After
diagonalizing the model~\cite{w2}, we construct the correlation
matrix reduced to {\it one} wire $a$, and then the $2N/M\times 2N/M$ matrix
${\mathbb A}$ (for simplicity we assume $N$ to be a multiple of $M$)
corresponding to the same wire.  After some algebra we obtain
\begin{eqnarray}
&&{\mathbb A}_{nm} = \Upsilon_a \sqrt{1-\Upsilon_a^2} \ {\mathbb B}_{nm}(1/2) 
\quad {\rm for}\;n\ne m,
\label{Emn2}\\
&&{\mathbb A}_{nn} = \Upsilon_a^2 \;\; {\rm for \; even\;}n,\quad 
{\mathbb A}_{nn} = (1-\Upsilon_a^2)  \;\; {\rm for \; odd\;}\,n,
\nonumber
\end{eqnarray}
where $\Upsilon_a^2 =(1 + {\mathbb S}_{aa})/2$.  Notice that ${\mathbb
A}$ has the same form of that for the two-wire problem,
cf. Eq.~(\ref{bmn2}), which is recovered setting
$\Upsilon_a=\epsilon/\sqrt{1+\epsilon^2}$.  Thus, using
Eq.~(\ref{asbehb}), we also derive the asymptotic behavior of the
entanglement entropy of any of the $M$ wires given by
Eq.~(\ref{asbehb}) with $s = 2 \Upsilon_a \sqrt{1 -
\Upsilon_a^2}=\sqrt{1-{\mathbb S}_{aa}^2}$.  Note that $s^2$ is the
total probability of a signal to be transmitted
from the wire $a$ to all other wires.  Being ${\cal C}_\alpha$ a
monotonous function of $s$, the maximum entanglement entropy is always
obtained for $s=1$ (${\mathbb S}_{aa}=0$), i.e., in the absence of
reflection.

{\it Non-equilibrium evolution}.  All results presented so far concern
the eigenstates of free-fermionic Hamiltonians.  However, the method
allows us to also obtain numerically exact and/or analytic
computations out of equilibrium.  Indeed Eq.~(\ref{cxy}) is valid if
we replace the one-particle eigenfunctions $\phi_k(x)$ with
appropriate solutions of the corresponding time-dependent one-particle
Schr\"odinger equation.  As an example that can be treated
analytically, we consider the off-equilibrium quantum dynamics of a
gas in a time-dependent harmonic potential.

Using the method outlined above and the knowledge of the eigenstates
of the harmonic oscillator, one can straightforwardly calculate the
ground-state (equilibrium) entanglement entropy of any spatial
subsystem $[x_1,x_2]$. In particular, the half-space entanglement
entropy, i.e., of $[-\infty,0]$, behaves as $S_\alpha(-\infty,0) =
\frac{1}{12}(1+\alpha^{-1})\ln N+ e_\alpha+O(N^{-{1\over\alpha}})$,
which can be analytically obtained, including the constant $e_\alpha$,
by developing results of Ref.~\cite{cv-10}.  In the presence of a
time-dependent potential $V(x)=\frac{1}{2} \kappa(t) x^2$, and
starting from the ground state for a given $\kappa_0=\kappa(t=0)$, we
can exploit the solution of the one-particle
problem~\cite{cv-10-2,GW-00} to infer that the time-dependent
entanglement entropy behaves as
\begin{equation}
S_{\alpha}(x_1,x_2;t)= S_{\alpha}(x_1/s(t),x_2/s(t);0)\,,
\label{sanxt}
\end{equation}
where $s(t)$ is an analytical function of the time-dependent potential
with $s(0)=1$. This shows the remarkable property that the evolution
in a harmonic potential simply corresponds to a global rescaling of
the system size.  In the case of a quantum quench with an
instantaneous removal of the harmonic potential of frequency
$\sqrt{\kappa_0}$, we have $s(t) =\sqrt{1+\kappa_0 t^2}$.  Then for
$t\to\infty$, $s(t)$ diverges and the entanglement of any finite
interval vanishes in the long time limit, while the entanglement of
any semi-infinite piece $[-\infty,x]$ tends asymptotically to
$S_{\alpha}(-\infty,0;0)$.

Other non-equilibrium situations such as local quantum quenches in
junctions of quantum wires (e.g., instantaneously turning on/off the
point contact at the vertex) can also be tackled within this
framework.  They may provide important insights in view of the recent
proposals of using the full counting statistics after a quench as an
experimental probe and a measure of entanglement~\cite{kl-08}.

{\it Discussions}.  In this Letter we have introduced a general
framework to calculate the entanglement entropy of gases that have a
representation in terms of free fermions.  The asymptotic entanglement
entropy grows like $\ln N$, with a prefactor that is given by the
central charge and that (in known cases) agrees with universal
predictions.  The UV regularization given by the number of particles
$N$ is alternative to the lattice spacing and, for the description of
gases, it has an immediate physical interpretation.  Moreover, the
formulas of the {\it universal features} of the entanglement are
easier to handle in the continuum at finite $N$ than on the lattice.

We conclude by mentioning some possible extensions of this work beyond
the straightforward (but still interesting) applications to other 1D
free gases.  Motivated by recent experiments in cold atoms, it is
interesting to consider 1D gases in an external random potential that,
in the case of a perfect gas, can be treated with our method.  All 
the results we have presented are in 1D, but the application of the method to
higher dimensions is straightforward.

\end{document}